\documentclass[12pt]{article}
\usepackage{graphicx}

\def\Title#1{\begin{center} {\Large #1 } \end{center}}
\def\Author#1{\begin{center}{ \sc #1} \end{center}}
\def\Address#1{\begin{center}{ \it #1} \end{center}}

\newenvironment{Abstract}{\begin{quotation}  }{\end{quotation}}

\begin{document}

\Title{\textbf{Twice More On Coulomb-Nuclear Interference}.}

\Author{ Vladimir A. Petrov\footnote {E-mail: {Vladimir.Petrov@ihep.ru}}  } 
\Address{ Logunov Institute for High Energy Physics, NRC Kurchatov Inst., Protvino, RF}

 \begin{flushright}
 "It's so hard to know what to do when 
 
 one wishes earnestly to do right."

Bernard Shaw.
 \end{flushright}

\begin{Abstract}
This is an improved and extended exposition of the modified formulas for accounting for Coulomb-nuclear interference in hadron scattering which is intended to dispel recently  claimed doubts.
\end{Abstract}

\def\thefootnote{\fnsymbol{footnote}}
\setcounter{footnote}{0}

\section{Introduction}
In paper \cite{Pet} we have analyzed the formula for accounting for Coulomb-nuclear interference (CNI) (approximation to the full amplitude $T_{C+N}(s,t)$ up to $ \mathcal{O}(\alpha) $ ) pushed forward in Ref.\cite{Ca} and \cite{Ku} and argued that although it is more correct than the Bethe formula used before it nonetheless  contained a superfluous term which stemmed from some oversimplifications when generalizing the case of point like charge to the realistic case of extended charges.
Moreover, in the same paper we gave an expression for the modulus of the full amplitude
$ T_{C+N} $ valid to all orders in $ \alpha $ expressed in terms of well defined functions without any kind of divergences, with physical (zero) photon mass and for an  arbitrary strong interaction amplitude.
More details on this subject were given in Ref.\cite{Pe} where alongside the exact formulas the perturbative expansion accounting for \textit{all} terms of order $ \mathcal{O}(\alpha^{2}) $ were derived. 

The modified formulas for account of CNI, being applied to the TOTEM data on the differential $ pp $ cross-section at 13 TeV, gave the values of $ \rho $  and $ \sigma_{tot} $ \cite{Ez} different from those published in \cite{TOT} and obtained with help of formulas and assumptions from Refs. \cite{Ca} and \cite{Ku}.

 In a recent paper \cite{Ka}, however, the formulas obtained in \cite{Pet}  were qualified as unreliable and were therefore subjected to a verification. Unfortunately, the very analytical derivation in question was left without attention and due critical examination. Instead there was chosen tracking the tendency of dependence on the decreasing parameter of fictitious "photon mass" $ \lambda $  down to the limits allowed by computer.
 
 The scheme of reasoning in Ref.\cite{Ka}  can be concisely summarized as follows:
  
 1. The concrete expression for $ \mid T_{C+N}(s,t)\mid^{2} $ from ref.\cite{Pet} which is the limit $ \lim _{\lambda}\rightarrow 0 \mid T_{C+N}(s,t;\lambda)\mid^{2} $, where 
 $ T_{C+N}(s,t;\lambda)$ stands for for the scattering amplitude with pseudo Coulombic exchanges by " photons" of non zero mass $ \lambda $, was considered in Ref.\cite{Ka} as obtained by means of "analytic manipulations" and "not well suited for numerical evaluation".
 
 2. Instead  the expression $ \mid T_{C+N}(s,t;\lambda) \mid^{2} $ was subjected (after several simplifications and assumptions of the corresponding integrands\footnote{See section 4 for our assessment of these considerations.} ) to numerical evaluation to show how fast the quantity
 \begin{equation}
 \frac{\mid T_{C+N}(s,t;\lambda) \mid^{2}}{\mid T_{C+N}(s,t)\mid^{2}} -1
 \end{equation}
 converges to zero with $ \lambda \rightarrow 0 $. With the aim to check if the explicit form of this limit obtained in Ref.\cite{Pet} is really achieved at $ \lambda \rightarrow 0 $.
 
 3. Due to some reasons the limit $ \lambda = 0 $ could not be reached, as the author of Ref.\cite{Ka} mentions. Instead he has chosen the  value $ \lambda = 3\cdot 10^{-5}  GeV $ as sufficiently well representing the zero. Also the subject of verification was taken not the full amplitude as treated in \cite{Pet} or \cite{Pe} but its approximation up to $ \mathcal{O}(\alpha ) $.
  The result was: the quantity
 \[\frac{\mid T_{C+N}(s,t) \mid^{2} (\mathcal{O}(\alpha) approximation from\cite{Pet}) }{\mid T_{C+N}(s,t;\lambda = 3\cdot 10^{-5} GeV) \mid^{2}( all \:orders \:in\: \alpha ) } -1\]
 deviates from zero at small $ -t < 0.01 $ up to $ 5\% $ while the $\mathcal{O}(\alpha)  $ approximation suggested in \cite{Ca} and \cite{Ku} does perfectly well and thus remains, according to the author of \cite{Ka},  the "best on the market". 
 
 This implies, "by default", a conclusion that the results of \cite{Pet},\cite{Pe}  should be qualified either wrong or useless (this concerns missing $ \mathcal{O}(\alpha^{2} ) $ terms).
 
 Paying all the tribute to the skill and ingenuity demonstrated in the work \cite{Ka}, we nevertheless do not consider the conclusion presented in Point 3 above convincing and our arguments are given in Section 4.
 
 To make things as clear as possible we believe it useful to once again demonstrate in all reasonable details our proof (to avoid imputations in "manipulations")of the formulas for accounting CNI so that everyone interested could detect our supposed mistakes and gaps, because, if to follow logically to the conclusions made in  \cite{Ka}, they must necessarily take place. 
 
We hope that experts will be condescending to the fact that throughout this article we will - in order to make the narration as clear and complete as possible - repeat many well-known facts, considerations and formulas.

 \section{General premises and formulation of the problem}
 
 In what follows  we will use our usual normalization when connecting amplitudes         (containing both Coulomb (C) and strong (N) interactions) in the momentum (transfer) $ q $ and in the impact parameter $ b $ spaces ( at the c.m.s. collision energy $ \sqrt{s} $) :
 \begin{equation}
 T_{C+N} (s, q) = 4s\int d^{2}b e^{iqb}  \tilde{T}_{C+N} (s, b).
 \end{equation}
  The amplitudes $ F (s,q) $ from Ref. \cite{ Ku, TOT, Ka}  are normalized as $ F = T/8\pi $.
  
 In the course of our derivations we will use 2D vector $ q $ which is related with the familiar invariants $ t $ and $ u $ as
 \[q^{2}= \frac{ut}{s-4m^{2}}\]
and we normally will not discern the designations $ q $ both for the vector and its modulus as in almost all the cases it is clear what is meant. If necessary, the transition from $ q $ to $t $ is accomplished without difficulty, so the use of both variables will be sometimes intermittent.

Integrals in $ q $  are taken over all $ q $-space because at the kinematic boundary $ q^{max} \approx \sqrt{s}/2 $ the amplitudes practically disappear at high energies and the difference is negligible. As a bonus, we can enjoy all the conveniences of using direct and inverse 2D Fourier transforms without vexing factors.
 
Due to the TOTEM/ALFA(ATLAS) measurements at the LHC  the proton-proton scattering is in the centre of interest and thus we will consider the $ pp $ scattering amplitude.
As one deals with the data for unpolarized protons with the cross-section
\[\frac{d\sigma}{dt}(s,t) = \frac{1}{16\pi s^{2}}\sum_{pol}\mid \mathcal{T}_{C+N}(s,q)\mid^{2}\]
where $ \mathcal{T}_{C+N}(s,q) $ is the $ pp $ scattering amplitude with definite spin projections and $ \sum_{pol} $ means both summation in final and averaging in initial
projections. From Fermi statistics it follows that 
\[\mathcal{T}_{C+N}(s,q)\] 
changes sign under the substitute $ t\leftrightarrow u $ and corresponding polarization indices changed  while the cross-section remains intact.

  As one deals with unpolarized scattering it is enough to model an effective scalar amplitude $ T_{C+N} $  so that
  \[\mid T_{C+N}\mid^{2} =  \sum_{pol}\mid \mathcal{T}_{C+N}(s,q)\mid^{2}. \]
Formally it remains an ambiguity which scalar amplitude to take : even or odd under 
$ t\leftrightarrow u $. Taking into account the effective character of modelling
we will choose the symmetric option : $ T(s,t,u)= T(s,u,t) $ because the antisymmetric amplitude would lead to exact zero of the differential cross-section at $ \theta_{c.m.s.} = 90^{\circ} $ which is not the case for the realistic case with account of polarizations.

Why this account of $ u-t $ symmetry is important? The matter is that when considering the expansion of the full $ C+N $ amplitude in powers of $ \alpha $ we  are faced with integration over transferred momenta and neglect of this circumstance may lead to an error. As was already said , the use of variable $ \textbf{q} $ does it automatically.
 
 To proceed further we use the standard eikonal representation
 \begin{equation}
 T_{C+N}(s, q) = -2is\int d^{2}b e^{iqb} [e^{2i\delta_{C+N}(s,b)}-1].
 \end{equation}
 
 Let us notice that in a real experimental set up the measured transferred momentum 
 $ q^{2} \approx -t $ is never exactly zero\footnote{The forward value of a generic observable $ \mathcal{F}(0) $ is always a product of one or another extrapolation procedure $ \lim_{t\rightarrow 0} \mathcal{F}(t)$.}  and therefore the contribution
of the second term in the integrand in Eq. (2), which is 
\[2is(2\pi)^{2} \delta^{(2)}(q),\]
vanishes if we keep $ q\neq 0 $ and we will always retain this condition except the cases when the amplitude is under integration.

So, we deal further on with the expression 
\begin{equation}
T_{C+N}(s, q) = -2is\int d^{2}b e^{iqb}e^{2i\delta_{C+N}(s,b)}
\end{equation}
and will not specially indicate the condition $  q\neq 0  $ for the "external" momentum transfer.
 
As all involved authors do, we assume the additivity of the eikonal (phase shift) w.r.t. Coulomb and strong interactions:
 \[\delta_{C+N}(s,b)= \delta_{C}(s,b) + \delta_{N}(s,b).\]
 Now, the Coulomb phase shift is assumed to be defined by the one-photon (Born) amplitude
 \begin{equation}
 \delta_{C}(s,b)= \frac{1}{4s}\int\frac{d^{2}q}{(2\pi)^{2}}e^{-iqb}T_{C}^{B}(s,q)
 \end{equation}

Taking into account that the bulk of the elastic $ pp  $ scattering is due to small angles $ \theta_{c.m.s.}, \pi - \theta_{c.m.s.} $ at which  $ t (u)\approx - q^{2}$ we will use the following approximation for the Coulomb Born amplitude (used, e.g., in Ref.\cite{Ca})
\begin{equation}
T_{C}^{B}(s,q)= -\frac{8\pi s \alpha F^{2}(q^{2}) }{q^{2}}.
\end{equation}
This expression is evidently $ t \leftrightarrow u $ invariant and approximates the exact values with a sufficient accuracy at relevant $ t $ and $ u $. A popular parametrization of the electromagnetic (Dirac) form factor is

\[F(q^{2})= (1+q^{2}/\Lambda^{2})^{-2}, \: \Lambda^{2} =0.71 GeV^{2}. \]

\section{Fictitious "photon mass" and how to get rid off it}
With the amplitude from Eq.(5) the integral in Eq.(4) diverges at small $ q $. To keep trace of this divergence one introduces a regulator, a fictitious "photon mass" $ \lambda $,  which changes the phase shift $\delta_{C}(s,b)  $ from Eq.(4) to the new one
\begin{equation}
\delta_{C}(b) \rightarrow \delta_{C}(b;\lambda)= - \alpha \int\frac{d^{2}q}{(2\pi)}e^{-iqb} \frac{F^{2}(q^{2})}{q^{2}+\lambda^{2}} = - \alpha \int_{0}^{\infty} \frac{qdq J_{0}(qb)F^{2}(q^{2})}{q^{2}+\lambda^{2}}.
\end{equation}
From now on we will omit the variable $ s $ from  all functions except its explicit appearing.

Thus, instead of the amplitude $ T_{C+N} (q)$ we have a regularized amplitude
\begin{equation}
T_{C+N} (q;\lambda) = -2is\int d^{2}b e^{iqb} e^{2i\delta_{C}(b;\lambda) + 2i\delta_{N}(b)}
\end{equation}
and the problem now is what to do with the fictitious mass and how , if we can, to come to the physical case $ \lambda = 0 $.

Let us consider the $ \lambda $ dependence of $ \delta_{C}(b;\lambda) $. 
As it is easy to see $ \delta_{C}(b;\lambda) $ can be represented in the form:

\[\delta_{C}(b;\lambda)= \delta_{C}(0;\lambda) + \alpha \int_{0}^{\infty} \frac{dq q}{q^{2}+ \lambda^{2}}(1-J_{0}(qb))F^{2}(q^{2})\doteq  \delta_{C}(0;\lambda) + \hat{\delta}_{C} (b;\lambda) \]
The first term, $\delta_{C}(0;\lambda)$, diverges at $ \lambda\rightarrow 0 $ (for definiteness we use the dipole parametrization of the form factor as is shown in the footnote):
\begin{equation}
\delta_{C}(0;\lambda)= \frac{\alpha}{2}\int _{0}^{\infty}\frac{dk^{2}}{k^{2}+\lambda^{2}}F^{2}(k^{2}) = \alpha \ln\frac{\Lambda}{\lambda} + const
\end{equation}
while the second one, $ \hat{\delta_{C}} (b;\lambda) $, remains finite\footnote{Such a method to eliminate the IR divergences is well known. See, e.g. Ref.\cite{La}.}. It is worth noting that such an extraction of the divergence is ambiguous: one can add to the diverging phase any quantity independent of $ b $ but this has no effect on the amplitude modulus. \textit{However, at the level of the very amplitude one can have different values.} Sure, the difference disappears upon passing to the modulus and physical photon mass. This is akin to the finite renormalization leaving the renormalized $ S $-matrix intact.

With all this in mind one can represent the regularized amplitude in the following form
\begin{equation}
T_{C+N} (q;\lambda)= e^{2i\alpha \delta_{C}(0;\lambda)} \cdot \hat{T}_{C+N} (q;\lambda)
\end{equation}
where
\[\hat{T}_{C+N} (q;\lambda)= -2is\int d^{2}b e^{iqb} e^{2i\hat{\delta}_{C}(b;\lambda) + 2i\delta_{N}(b)}\]

Now we see that 

\[\mid T_{C+N} (q;\lambda) \mid^{2} = \mid \hat{T}_{C+N} (q;\lambda)\mid^{2}\]
and we have good news: we can pass to the physical limit ( for which we justifiably use the same designation as in Eq.(3) because, as we see, the modulus of $ T_{C+N} $ is IR finite)
\[\lim_{\lambda \rightarrow 0}\mid T_{C+N} (q;\lambda) \mid^{2}\doteq \mid T_{C+N} (q) \mid^{2} \]
which is safe and 
\begin{equation}
\mid T_{C+N} (q) \mid^{2}=\mid \hat{T}_{C+N} (q;0)\mid^{2}
\end{equation}
where the eikonal has the form
\begin{equation}
\delta_{C+N}(b) = \delta_{C}(b)+ \delta_{N}(b) = \alpha \int_{0}^{\infty} \frac{dk}{k} F^{2}(k^{2})(1-J_{0}(kb)) + \delta_{N}(b)
\end{equation}
 We use the same designation $ \delta_{C}(b)$ for the first term in r.h.s of Eq.(4) as  for "non-renormalized" phase in Eq.(4). We hope it will not cause misunderstanding.
\section{"No regularization" argument}
  
It is instructive to show that for the modulus of the amplitude $ T_{C+N} (q) $ we do not need to use any dummy "masses" at all from the very beginning.

In fact, let us write down the modulus squared of the amplitude $ T_{C+N} (q) $  as in Eq.(2). With additive eikonal $\delta_{C+N}(b)  $ we get 
\begin{equation}
\mid T_{C+N}\mid^{2} (q) = 4s^{2}\int d^{2}b^{'}d^{2}b^{''}e^{iq(b^{'}-b^{''})}e^{2i\alpha \Delta_{C} (b^{'},b^{''})} e^{2i[\delta_{N}(b^{'})-\delta^{\ast}_{N}(b^{''})]}
\end{equation}
where
\[\Delta_{C} (b^{'},b^{''})= \int_{0}^{\infty} \frac{dk}{k}F^{2}(k^{2})[J_{0}(b^{''}k) - J_{0}(b^{'}k)].\]
 The integral in the second line  converges both at small and large $ k $. Also the integration in Eq.(13) in impact parameters exists not only as a distribution but as a continuous function of $ q^{2} $ at real $ q^{2} > 0 $.
In expanded form Eq.(13) looks as follows:
 \begin{equation}
\mid T_{C+N}\mid_{q\neq0}^{2} = 4s^{2} S^{C} (q,q) + \int\frac{d^{2}q^{'}}{(2\pi)^{2}}\frac{d^{2}q^{''}}{(2\pi)^{2}} S^{C} (q^{'},q^{''})T_{N} (q-q^{'})T_{N}^{\ast} (q-q^{''})
\end{equation}
\[+4s \int\frac{d^{2}q^{'}}{(2\pi)^{2}} Im[S^{C} (q,q^{'})T_{N}^{\ast} (q-q^{'})]\]
where
\begin{equation}
S^{C} (q^{'},q^{''})= \int d^{2}b^{'}d^{2}b^{''} e^{i{q}^{'}{b}^{'}-i{q}^{''}{b}^{''}} e^{2i\alpha \Delta_{C} ( b^{'},\, b^{''})}.
\end{equation}

  For an explicit illustration, in the case of the point like electric charges ( $ F=1 $ ) we would get 
 \[\Delta_{C,\: point} (b^{'},b^{''}) = ln\frac{b^{'}}{b^{''}};\;\]
and
 \begin{equation}
S^{C} (q^{'},q^{''}) =  (4\pi\alpha)^{2} \frac{(q^{''2}/q^{'2})^{i\alpha}}{q^{'2}q^{''2}}.
\end{equation}
When obtaining these expressions no IR regularization was needed. Evidently, the expression (13)(or (14)) coincides with $ \mid \hat{T}_{C+N}(q;0)\mid^{2} $ from Eq.(10).

Thus we have proved that the modulus of the scattering amplitude with account of both Coulomb and strong interactions exists for the physical (zero) photon mass while this is not the case for its phase which diverges like $ \ln \lambda $.
So when considering the observable quantity
\begin{equation}
\frac{d\sigma}{dt} = \frac{1}{16\pi s^{2}}\mid T_{C+N} (q) \mid^{2}= \frac{1}{4\pi} \mid\int d^{2}b e^{iqb} e^{2i\delta_{C}(b)+2i\delta_{N}(b)} \mid^{2}
\end{equation}
one, in principle, could well do (with use of Eq.(11)) without resorting to unobservable entities like "fictitious masses" or else. 

From the common sense it is clear that no "numerical check"  can "prove" or "disprove" the formulas (13)-(14). Numerical integration can only give numerical estimates of the expression (14) for one or another choice of functions $ \delta_{N} $ and compare the result with the data.

We would like to emphasize, once again, that this statement is not the author original invention and is well known. We repeat it just because it appears that there is still some ignorance of this subject.

Now let us come back to the conclusions made in \cite{Ka} about numerical verification of the above said results.

To better adapt integration in the impact parameter to computer integration 
an effective cut-off is defined in \cite{Ka} which is to specify the limits $ b_{max} $ of actual numerical integration. The asymptotic value (as follows from Eq.(7) $ \delta_{C}(b;\lambda)\mid_{b\rightarrow\infty} \approx K_{0}(\lambda b) $ with $ K_{0} $ the MacDonald cylindrical function of the zero order)
\begin{equation}
\delta_{C}(b;\lambda)\mid_{b\rightarrow\infty} \sim e^{-\lambda b}
\end{equation}
was used and it was 
assumed on this ground that the impact parameters $ b $ obeying the inequality
\[e^{-\lambda b} \geq e^{-c}\]
with $ c=10 $ practically saturate the integration in question with
\[b_{max} = \frac{c}{\lambda}.\]

 However, as we said already more than once, the modulus of the physical amplitude $ \mid T_{C+N}\mid $
has nothing to do with fictitious parameters like $ \lambda $ which only influence the \textit{unobservable} and ambiguous overall phase (e.g. $ \delta_{C} (0;\lambda)  $ ).

To make the point more explicit let us consider the large impact parameter behaviour of the integrand in Eq.(13)which is defined at large impact parameters by the factor (the second line in Eq.(15))
\[exp(2i[\delta_{C}(b^{'})- \delta_{C}(b^{''})]) \approx (b^{'2}/b^{''2})^{i\alpha}.\]
The contribution of the form factors $\sim exp(i\alpha (exp(-\Lambda b^{'})-exp(-\Lambda b^{''})$) and may be neglected.

We see that the effective cut-off both in $ b^{'} $ and $ b^{''} $ is rather $ 1/q $ than something else. Indeed, let us take for illustration the point like case ($ F=1 $).  The leading term in $ \mid T_{C+N}\mid^{2} $ at very small $ q $ is certainly $ \sim 1/q^{4} $ which perfectly agrees with the above said. 
From this viewpoint the integration limits assumed in \cite{Ka} do not look justified.

Thus, it seems that numerical arguments presented in \cite{Ka} as a basis for refutation of our results \cite{Pet} and \cite{Pe} have no solid ground.
It is difficult to understand why it seemed impossible or inconvenient, as mentioned in \cite{Ka} , to numerically integrate directly the physical expression (13) for $ \mid T_{C+N}\mid $ which is even simpler than that with an artificial parameter $ \lambda $.

\section{Conclusion}
 In this note we have considered an objection pushed forward in \cite{Ka} against our paper \cite{Pet} (see also \cite{Pe}) where new formulas to account for CNI were deduced. 
 We presented an explicit and a more detailed derivation of the results given in \cite{Pet}
 to enable any reader to verify the validity of our arguments and reasoning.
 
 We have shown that the arguments based on use of fictitious parameters for estimation of the observable quantities (independent of such parameters by definition) are inconclusive.

The use of  formulas for CNI as given in \cite{Pet} (see also
 \cite{Pe}) lead to new values of such important parameters as $ \rho $ and $ \sigma_{tot} $ \cite{Ez} different  from those which were presented in \cite{TOT}.
 
\section{Acknowledgements}

This work was inspired by the correspondence with Jan Ka\v{s}par who acquainted the author with a preprint of his work \cite{Ka} before making it public.  I am very much grateful to him.

 I am also thankful to Vladimir Ezhela, Nikolay Tkachenko and Anatoliy Samokhin for useful discussions.

\end{document}